\documentclass{elsart}
\usepackage{amscd,graphicx,psfrag,bm,amsmath,amssymb,natbib}

\newcommand{\hvect}[1]{\mathbf{#1}}
\newcommand{\vect}[1]{\mathbf{#1}}
\newcommand{\vectsym}[1]{\boldsymbol{#1}}
\newcommand{\tensor}[1]{\bm{#1}}

\renewcommand{\exp}[1]{\text{e}^{#1}}

\newcommand{\avg}[1]{\left\langle #1 \right\rangle}
\renewcommand{\r}{\vect{r}}

\newcommand{\stressAV}{\langle \sigma_{xx}\rangle}
\newcommand{\sigmaY}{\sigma_{\rm Y}}
\newcommand{\sigmaN}{\sigma_{\rm N}}
\newcommand{\epsY}{\varepsilon_{\rm Y}}

\newcommand{\tauY}{\tau_{\rm Y}}
\def\micron{\:\mu\mbox{m}}

\begin{document}
\begin{frontmatter}
\title{Study of size effects in thin films by means of a crystal plasticity
  theory based on DiFT}
\author[1]{S. Limkumnerd}
\ead{s.limkumnerd@rug.nl}
\author[1]{E.~Van der Giessen\corauthref{cor}}
\corauth[cor]{Tel: +31-50-3638046; Fax: +31-50-3634886;}
\ead{E.van.der.Giessen@rug.nl} 

\address[1]{Zernike Institute for Advanced Materials, University of
  Groningen, Nijenborgh 4, NL-9747 AG Groningen, The Netherlands}

\begin{abstract}

In a recent publication, we derived the mesoscale continuum theory of
plasticity for multiple-slip systems of parallel edge dislocations,
motivated by the statistical-based nonlocal continuum crystal
plasticity theory for single-glide due to \cite{YefiGromGies04}. In
this dislocation field theory (DiFT) the transport equations for both
the total dislocation densities and geometrically necessary
dislocation densities on each slip system were obtained from the
Peach--Koehler interactions through both single and pair dislocation
correlations. The effect of pair correlation interactions manifested
itself in the form of a back stress in addition to the external shear
and the self-consistent internal stress. We here present the study of
size effects in single crystalline thin films with symmetric double
slip using the novel continuum theory. Two boundary value problems are
analyzed: (1) stress relaxation in thin films on substrates subject to
thermal loading, and (2) simple shear in constrained films. In these
problems, earlier discrete dislocation simulations had shown that size effects are born out of layers of dislocations developing near constrained interfaces. These boundary layers depend on slip orientations and applied loading but are insensitive to the film thickness. We investigate stress response to changes in controlled parameters in both problems. Comparisons with previous discrete dislocation simulations are discussed.

\end{abstract}


\end{frontmatter}

\section{Introduction}
Contrary to the prediction of classical crystal plasticity theory,
experimental observations at length scales ranging from hundreds of
nanometers to tens of microns show size effects of the type ``smaller
is
harder''~\citep{EbelAshb66,BrowHam71,FlecMullAshbHutc94,MaClar95,StolEvan98,Arzt98}.
This failure of conventional continuum theory is caused by the lack of a characteristic length scale. Several more sophisticated theories~\citep{Aifa84b,WalgAifa85,FlecHutc93,FlecMullAshbHutc94,OrtiRepe99,OrtiRepeStai00,AchaBass00,AchaBeau00,BassNeedGies01,Gurt00,Gurt02,Gurt03} have been developed which attempt to incorporate a length scale through the concept of geometrically necessary dislocations (GNDs) as introduced by~\cite{Nye53}. In these theories, however, the length scale enters in an ad-hoc fashion, and often has to be supplied a priori by comparison with discrete dislocation simulations or experimental results.

Alternatively, \cite{YefiGromGies04,YefiGromGies04b} have applied a
nonlocal continuum plasticity theory based on work by \cite{Grom97}
and \cite{ZaisMiguGrom01} to successfully solve a set of
boundary-value problems for systems with one active slip
system.\footnote{Somewhat similar approaches have been taken by
\cite{ArsePark02,ArseParkBeck04,ElAz00,LimkSeth06,AchaRoy06,RoyAcha06}.}
They described the evolution of total dislocation densities and GND
densities using a set of coupled transport equations. In addition to
external shear and Peach--Koehler interactions among dislocations, the
effect of pair-dislocation correlation, in the form of a back stress,
was considered; the latter gave rise to a natural length scale
$1/\sqrt{\rho}$, determined
by the average dislocation spacing $\rho$. Thriving on the success of
their theory, \cite{YefiGies05,YefiGies05b} attempted to extend their
single-slip theory to describe multiple-slip systems on
phenomenological grounds. Albeit favorable results were obtained in
the problem of shearing of thin films, the theory could not capture
the size and orientation dependent hardness observed in thin films.

To address this problem, we have reformulated the multiple-slip theory aiming to extract the correct angular dependence of the back stress between different pairs of slip orientations~\citep{LimkGies07a}. By solving Bogolyubov--Born--Green--Yvon--Kirkwood (BBGYK) integral equations that relate different orders of dislocation correlation functions, the functional forms of pair-dislocation densities were derived. The results provided slip-orientation dependence of pair densities from which the exact expression of the back stress was obtained. In their recent publication, \cite{GromGyorKocs06} arrived at the same expression for a pair correlation function in the case of single-slip systems.

We begin in Sec.~\ref{S:DFT} by giving a summary of our continuum
theory with a short account to the work of \cite{YefiGies05}. In
Sec.~\ref{S:ThinFilm}, we apply the theory to the problem of stress
relaxation in single crystalline thin films on substrates subjected to
thermal loading. It was this problem in which the results between the
former multiple-slip theory \citep{YefiGies05b} and discrete
dislocation simulations \citep{NicoGiesNeed03,NicoGiesNeed05} deviated
most. In a quasi-static limit, where dislocations rearrange themselves
much faster than the stress increase in the film, an analytical
solution is derived. The hardening effect due to the film thickness
and comparisons with the discrete dislocation results can be directly
investigated for two slip orientations.
Finally in Sec.~\ref{S:Shearing}, we revisit the problem of the simple
shear response of thin films, which was used by \cite{YefiGies05} for selecting their slip-interaction law. Layers of dislocations form on the top and bottom boundaries which give rise to size effects. Analytical solutions of our theory are checked against the discrete dislocation simulations by \cite{ShuFlecGiesNeed01}.

\section{Summary of DiFT-based plasticity}\label{S:DFT}
Over a decade ago, \cite{Grom97} has derived a set of transport
equations governing the motion of many-dislocation densities by
carrying out a statistical averaging procedure on ensembles of
edge dislocations on parallel glide planes. \cite{ZaisMiguGrom01}
later on specialized these equations to describe evolution of
single-dislocation densities in terms of pair-dislocation
densities. Recently the authors have extended the above formalism to
include systems with more than one active
slips~\citep{LimkGies07a}. By constructing the integral equations that
relate different orders of dislocation correlation functions, we
explicitly calculate pair correlation functions, and hence
pair-dislocation densities. In this section we shall briefly summarize
this continuum theory, leading the derivation to \cite{LimkGies07a}.

Consider a single crystal with $N$ active slip systems where each system $i$ is defined by slip direction $\hvect{s}_i$ and slip plane normal $\hvect{m}_i$. We assume that the motion of dislocations is overdamped; positive dislocations on slip system $i$ flow with velocity $\vect{v}_i \equiv (\vect{b}_i/B)\,\tau^\text{eff}_i$ in the direction of their Burgers vector $\vect{b}_i \equiv b\,\hvect{s}_i$, with magnitude proportional to effective resolved shear stress $\tau^\text{eff}_i$ with drag coefficient $B$, while negative dislocations flow in the opposite direction. The evolution equations for uncorrelated, single-dislocation densities $\rho^+_{i}$ and $\rho^-_{i}$ can then be re-written in terms of a set of coupled transport equations for total dislocation density $\rho_i \equiv \rho^+_i+\rho^-_i$ and the GND density $\kappa_i \equiv \rho^+_i-\rho^-_i$ as follows:
\begin{equation}\label{E:rho_kappa_evol2}
\partial_t \rho_i + \vectsym{\nabla}\cdot\left[ \kappa_i \vect{v}_i \right] = 0,\qquad
\partial_t \kappa_i + \vectsym{\nabla}\cdot\left[
\rho_i \vect{v}_i \right] = 0, \mbox{ (no sum over $i$)}
\end{equation}
with $\vectsym{\nabla}$ the derivative with respect to spatial position
$\r$. Nucleation and annihilation of dislocations can be taken into account by
modifying the right-hand side of the evolution law for $\rho_i$
(cf.~\cite{YefiGromGies04}). The dislocation density description can
be incorporated into the framework of crystal plasticity through Orowan's relation
\begin{equation}\label{E:Orowan}
	\dot{\gamma}_i = \frac{b^2}{B}\,\tau^\text{eff}_i \rho_i 
\end{equation}
and the definition of plastic strain rate
\[
\dot{\tensor{\varepsilon}}^\text{p} \equiv \sum_{i=1}^N 
     \dot{\gamma}_i \tensor{P}_i \,,\quad
\tensor{P}_i = \half(\vect{s}_i \otimes \vect{m}_i 
                   + \vect{m}_i \otimes \vect{s}_i) \,.
\]
Substitution into the second dynamical equation in
(\ref{E:rho_kappa_evol2}) and time integration yields
Kr{\"o}ner's relation
\begin{equation}\label{E:Kroner}
	\kappa_i = -(1/b)(\hvect{s}_i\cdot\vectsym{\nabla})\gamma_i\,,
\end{equation}
which connects GND density $\kappa_i$ to plastic slip $\gamma_i$. 

The effective resolved shear stress
\begin{equation}\label{E:tauEffective}
 	\tau^\text{eff}_i \equiv \tau_i -\tau_i^\text{b},
\end{equation}
consists of $\tau_i$---the external shear stress plus the self-consistent,
long-range, single-dislocation interaction---and the back stress
$\tau_i^\text{b}$ given by
\begin{equation}\label{E:taub}
 	\tau_i^\text{b}(\r) = \frac{\mu b D}{2\pi(1-\nu)} \sum_{j=1}^N \cos(\theta_{ij}) \frac{(\vect{b}_j\cdot\vectsym{\nabla})\kappa_j(\r)}{\rho_j(\r)},
\end{equation}
arising from the short-range, dislocation-dislocation interactions. Here $\mu$
and $\nu$ are the shear modulus and the Poisson ratio respectively. The
strength of intra-slip back stress is controlled by the dimensionless constant
$D$. The back stress contribution from slip system $j$ to slip system $i$ is
reduced relative to the self back stress by a factor $\cos(\theta_{ij})$, where $\theta_{ij}$ is the angle between planes of slip system $i$ relative to $j$.

The form of the back stress as shown in Eq.~(\ref{E:taub}) reduces to that of
the previous single-slip
theory~\citep{GromCsikZais03,YefiGromGies04,YefiGromGies04b} for $N=1$. The
$\cos(\theta_{ij})$ slip-interaction coupling considered here also appears in
the strain-gradient theory for continuum crystal plasticity by
Gurtin~\citep{Gurt00,Gurt02,Gurt03}. In an early attempt to extend their
theory to describe systems with multiple slips, \cite{YefiGies05} had
considered three different coupling terms: $\cos^2(\theta_{ij})$,
$\cos(2\theta_{ij})$, and $\cos(\theta_{ij})$. They subsequently discarded the
first and the third variations upon comparisons with discrete dislocation
simulations by \cite{ShuFlecGiesNeed01}. Although the chosen form of coupling
showed reasonable agreements with the discrete dislocation results in the
problem of simple shearing of constrained thin film, it failed to capture the
dependence on film size and slip-orientation observed in the problem of stress
relaxation in thin films on substrates~\citep{YefiGies05b}. We shall reexamine
these problems with the new continuum theory in the following sections and
argue that the success of the $\cos(2\theta_{ij})$-type coupling was just fortuitous.

\section{Application to single crystal thin films on a substrate}\label{S:ThinFilm}
\begin{figure}[htb]
\psfrag{s1}{$\hvect{s}_1$}
\psfrag{m1}{$\hvect{m}_1$}
\psfrag{s2}{$\hvect{s}_2$}
\psfrag{m2}{$\hvect{m}_2$}
\psfrag{f}{$\phi$}
\psfrag{y}{$\hvect{y}$}
\psfrag{af}{$\alpha_f$}
\psfrag{as}{$\alpha_s$}
\psfrag{H}{$H$}
\includegraphics[width=0.8\linewidth]{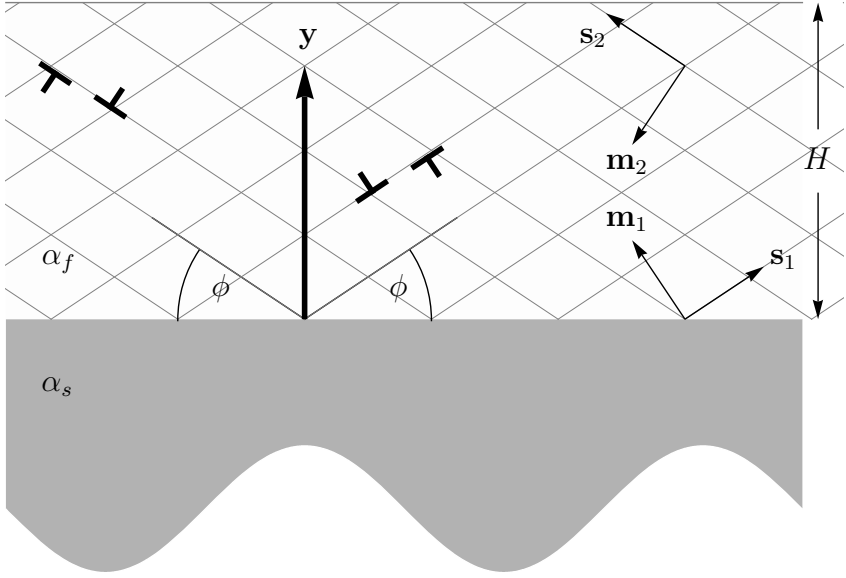}
\caption{\label{fig:ThinFilm}A thin film of thickness $H$ and thermal expansion coefficient $\alpha_f$ is situated on top of an infinite substrate with coefficient $\alpha_s$. The film has two symmetrical slip planes defined by angle $\phi$. The $\hvect{y}$ axis is taken to be perpendicular to the film--substrate interface.}
\end{figure}

In this section we consider the problem of stress relaxation in a single crystalline thin film, oriented
for symmetric double slip, on a substrate subjected to thermal loading. 
The geometry of the problem is shown in Fig.~\ref{fig:ThinFilm}. Initially
both the thin film, with thermal expansion coefficient $\alpha_f$, and the
substrate, with coefficient $\alpha_s$, are at a (high) temperature $T_0$. Since
$\alpha_f >\alpha_s$, a tensile stress up in the film as temperature
decreases (with a rate $\dot T$). At sufficiently high stress, pairs of
dislocations nucleate on the two active slips according to Frank--Read
mechanism. When the material is assumed to be initially homogeneous, the problem is effectively
one-dimensional; only variations along the direction perpendicular to the film
matter and the only non-vanishing stress component is
$\sigma_{xx}\equiv\sigma$. Also, by symmetry, $\gamma_1=-\gamma_2$ and 
$\tau_1=-\tau_2$. Hence, 
on average, the density of positive dislocations on the first slip is
the same as that of negative dislocations on the second, while the negative of
the first slip and the positive of the second are driven out of the system
through the top traction-free surface. We
shall henceforth drop the subscripts and only consider slip system 1. 

This problem can be treated rather simply in a quasi-static limit where dislocations rearrange
themselves much faster than the stress change. In this limit, the exact
expressions for nucleation and/or annihilation terms are unimportant and the
nature of the evolution equation (\ref{E:rho_kappa_evol2}) is only to
transport dislocations inside the thin film according to its overall
effective stress.  At any particular time, the distribution of these
densities can be calculated from the competition between the back
stress and the stress due to the thermal mismatch. Given the form of
the back stress (\ref{E:taub}), we can derive the time-dependence of
this expression from the compatibility requirement in terms of slip
$\gamma_i$ on system 1 and 2. Using Kr\"oner's relation
(\ref{E:Kroner}), the time evolution of the overall resolved shear
stress as a function of slip orientation can then be found. The
effects of film thickness and slip orientation on the stress response can be investigated from these expressions.

In the absence of plasticity, the stress inside the film would build up according to
\begin{equation}
	\sigmaN(T) = 2\mu\left(\frac{1+\nu}{1-\nu}\right)\alpha(T_0 -T)\,,
\end{equation}
where $\alpha \equiv \alpha_f - \alpha_s$ is the effective expansion
coefficient of the film relative to the substrate. Once the yield
point is reached, $\sigmaY \equiv \sigmaN(T_{\rm Y})$, plastic
straining,
\[
	\dot{\varepsilon}_{xx} = - \dot{\gamma}\sin(2\phi) ,
\]
is governed by the resolved shear stress
\[
 	\tau = -\frac{\sin(2\phi)}{2} \sigma .
\]
Compatibility of the thermally-induced strain and the elastoplastic strains
requires that after the yield point, is reached,
\begin{equation}\label{E:eq1}
	(1+\nu)\alpha \Delta T = \frac{1-\nu}{2\mu}\,(\sigma-\sigma_{\rm Y}) - \gamma\sin(2\phi)
\end{equation}
where $\Delta T \equiv T_{\rm Y}-T$ is the temperature drop since yield, and
$\gamma$ is the plastic slip (taken to be of slip system 1).
The effective shear stress $\tau^{\rm eff}$ comprises the resolved shear stress
\begin{equation}\label{E:eq2}
 	\tau = -\frac{\sin(2\phi)}{2}\left[\sigmaN + \frac{2\mu}{1-\nu}\,\gamma\sin(2\phi) \right]\!,
\end{equation}
and the back stress which, according to (\ref{E:taub}), is given by
\begin{equation}\label{E:eq3}
\begin{split}
 	\tau^{\rm b} &= \frac{\mu b D}{2\pi(1-\nu)\rho}\sin(\phi)\left[1-\cos(2\phi)\right]\partial_y \kappa \\
	&= \frac{\mu bD\sin^3(\phi)}{\pi(1-\nu)\kappa}\,\partial_y \kappa\,,
\end{split}
\end{equation}
since $\rho = \kappa$ in this system.
Combining eqs. (\ref{E:eq1})--(\ref{E:eq3}) with 
$\kappa=-\sin(\phi)/b {\partial_y \gamma}$ from (\ref{E:Kroner}), 
we can write the
effective shear stress as:
\begin{equation}\label{E:tauThinFilm}
 	\tau^{\rm eff} = -\frac{\sin(2\phi)}{2}\frac{2\mu}{1-\nu}\bigg[\frac{1-\nu}{2\mu}\,\sigmaN  \\ +\gamma\sin(2\phi) + \frac{bD}{\pi}\frac{\sin^3(\phi)}{\sin(2\phi)}\,\frac{\partial^2_y \gamma}{\partial_y \gamma} \bigg]
\end{equation}

Under the quasi-static assumption mentioned above, the equation of motion
(\ref{E:rho_kappa_evol2}) is solved by force balancing---in other words---by
setting $\tau^{\rm eff}=0$. Eq.~(\ref{E:tauThinFilm}) then gives the nonlinear differential equation
\begin{equation}\label{E:nonlinearODE}
 	\frac{\zeta f(\phi)}{2}\, \partial_y^2\gamma + \partial_y\gamma \left[\gamma\sin(2\phi) + \frac{1-\nu}{2\mu}\,\sigmaN\right] = 0
\end{equation}
with the length scale $\zeta \equiv 2bD/\pi$ being considered the new fitting parameter (instead of $D$). 
The solution during yield, subject to the no-slip condition $\gamma = 0$ at the film--substrate interface $y=0$, is unique and given by
\begin{subequations}
\begin{align}
  	\sin(2\phi)\,\gamma &= -\frac{1-\nu}{2\mu}\left[\sigmaN - \sigmaY\,\frac{\sigmaN\cosh(\lambda y) +\sigmaY\sinh(\lambda y)}{\sigmaY\cosh(\lambda y)+\sigmaN\sinh(\lambda y)} \right]\!, \\
	\lambda &\equiv \frac{\epsY}{f(\phi)}\frac{1}{\zeta}\,.
\end{align}
\end{subequations}
Here, $\epsY = (1+\nu)\alpha (T_0-T_{\rm Y})$ is the film's
strain at yield, and $f(\phi)\equiv \sin^3(\phi)/\sin(2\phi)$ contains the angular dependence on slip orientation. The stress profile after yield can be derived using Eq.~(\ref{E:eq2}):
\begin{equation}\label{E:filmstress}
 	\sigma(y,T) = \sigmaY\,\frac{\sigmaN\cosh(\lambda y) +\sigmaY\sinh(\lambda y)}{\sigmaY\cosh(\lambda y)+\sigmaN\sinh(\lambda y)}
\end{equation}
The average stress over the thickness of the film, $\stressAV \equiv
(1/H)\int_0^H \sigma\,dy$, follows directly from
Eq.~(\ref{E:filmstress}) as
\begin{equation}\label{E:stressAV}
 	\langle \sigma(T)\rangle = \frac{\sigmaY}{\lambda H}\log\!\left[\cosh(\lambda H)) + \frac{\sigmaN(T)}{\sigmaY}\,\sinh(\lambda H) \right]
\end{equation}

To compare results between the non-local theory and discrete
dislocation simulations by \cite{NicoGiesNeed03} we use parameters
from their simulation. The film is taken to be isotropic with
Poisson's ratio $\nu = 0.33$, Young modulus $E = 70$~GPa (from which
the value of $\mu = E/(2(1+\nu))$ is computed), and thermal expansion
coefficient $\alpha_f = 23.2\times 10^{-6}\mbox{K}^{-1}$. These values
are representative of aluminum. The silicon substrate has expansion
coefficient $\alpha_s = 4.2\times 10^{-6}\mbox{K}^{-1}$. The system is
cooled from an initial temperature of $T_0 = 600$~K down to $T =
400$~K, at a rate of $\dot T = 4\times 10^7$~K/s. For the
source density and source strength (distribution) chosen by
\cite{NicoGiesNeed03}, yield starts when the temperature reaches
$T_{\rm Y} \simeq 582$~K, i.e. at 
$\sigma_{\rm Y} \simeq 35.8$~MPa.

\begin{figure}[htb]
\psfrag{stress}{$\sigma$ (MPa)}
\psfrag{thickness}{$y$ ($\mu$m)}
\psfrag{Honemicron}{$H = 1.0~\mu$m}
\psfrag{pfive}{$0.5~\mu$m}
\psfrag{ptwofive}{$0.25~\mu$m}
\includegraphics[width=0.8\linewidth]{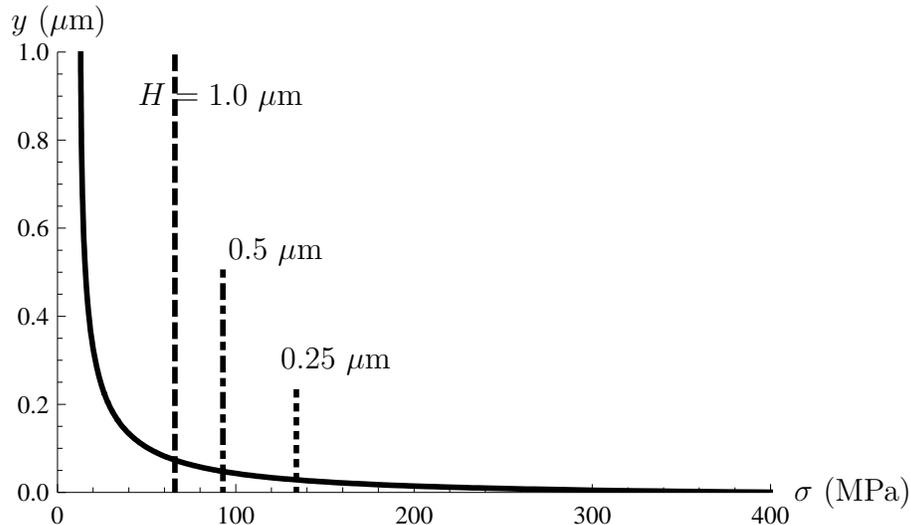}
\caption{\label{fig:StressProfile}Stress distribution across the film as predicted by Eq.~(\ref{E:filmstress}) for $\phi=60^\circ$. Vertical lines indicate the average stress in the films at different film thicknesses $H$.}
\end{figure}
Fitting to the average film stress at $T=400$~K predicted by the discrete dislocation simulations for orientation $\phi = 60^\circ$ yields a value of $\zeta \simeq 28.5$~nm.  
Fig.~\ref{fig:StressProfile} shows the corresponding stress
distribution across the film thickness according to
Eq.~(\ref{E:filmstress}). At the film--substrate interface, the stress
$\sigma$ reaches its elastic value of $\sigmaN \simeq 397$~MPa, and
decays roughly exponentially to the yield stress $\sigmaY \simeq
35.8$~MPa at the free surface. This profile is independent of the film
thickness $H$, as is the discrete dislocation result for the thickest
two films. The average stress $\stressAV$ for each thickness is
indicated by a vertical line. The result for $\phi = 30^\circ$
exhibits a similar functional dependence but with a steeper decay, and is omitted for brevity.

\begin{figure}[htb]
\centering
\psfrag{stress}{$\stressAV$ (MPa)}
\psfrag{temp}{$T$ (K)}
\psfrag{onemic}{$1.0~\mu$m}
\psfrag{pfivemic}{$0.5~\mu$m}
\psfrag{Hptwofivemic}{$H = 0.25~\mu$m}
\includegraphics[width=0.8\linewidth]{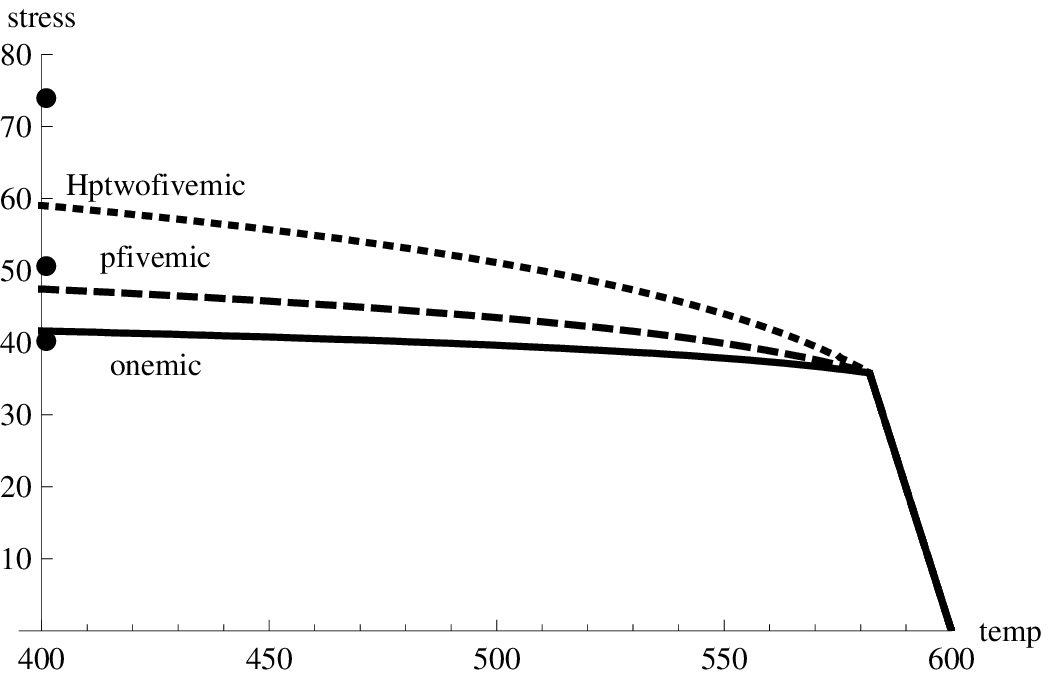}
(a)
\includegraphics[width=0.8\linewidth]{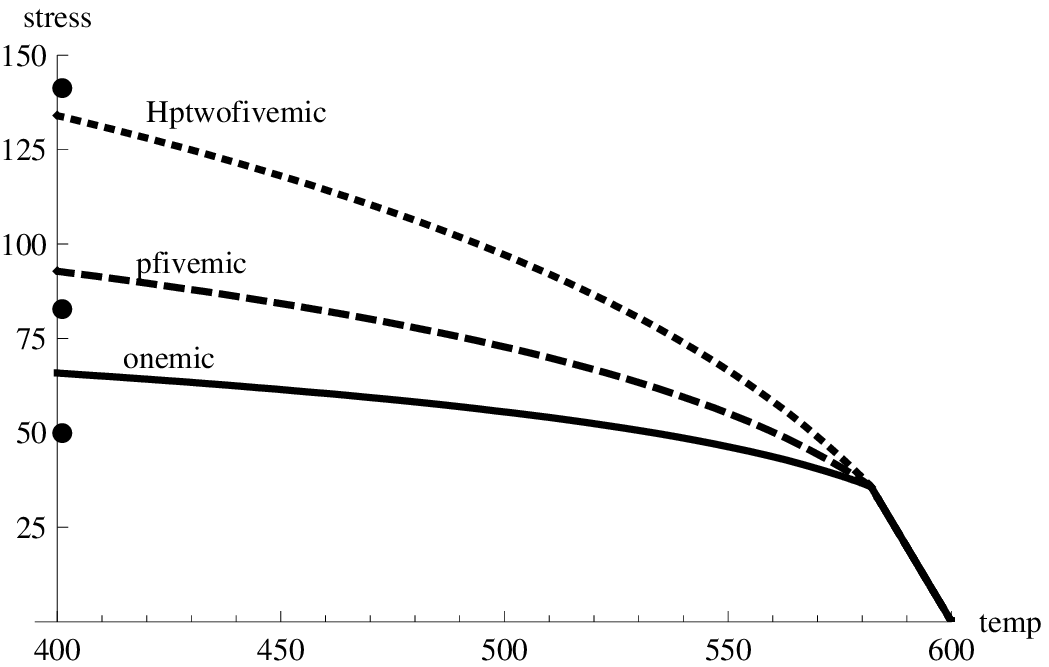}
(b)
\caption{\label{fig:StressTemp}Temperature dependence of average tensile
  stress $\stressAV$ from Eq.~(\ref{E:stressAV}) for slip orientations (a)
  $\phi = 30^\circ$ and (b) $\phi = 60^\circ$. The solid dots represent the
  discrete simulation results \citep{NicoGiesNeed03}}
\end{figure}

\begin{figure}[hbt]
\psfrag{stress}{$\stressAV$ (MPa)}
\psfrag{thickness}{$H$ ($\mu$m)}
\psfrag{thirty}{$\phi = 30^\circ$}
\psfrag{sixty}{$\phi = 60^\circ$}
\includegraphics[width=0.9\linewidth]{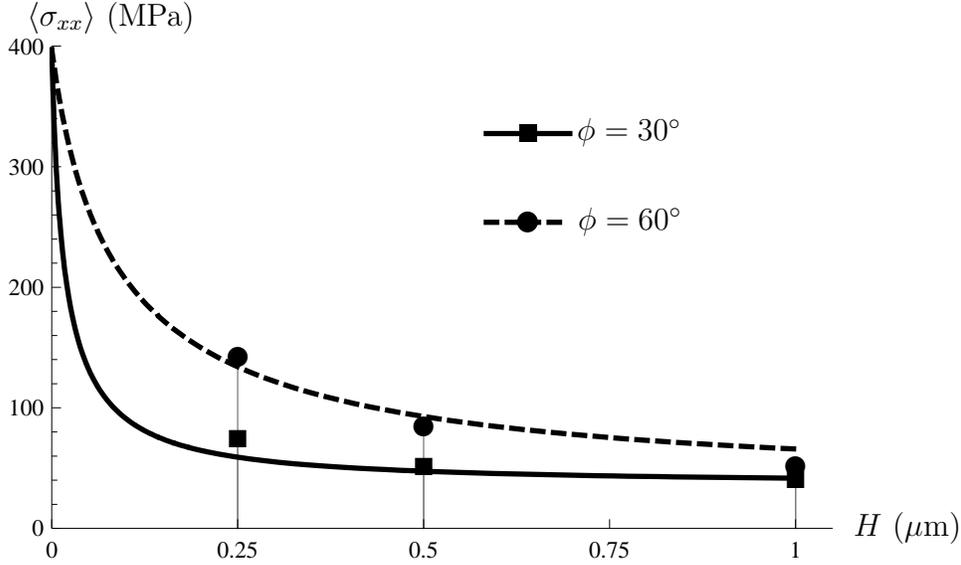}
\caption{\label{fig:StressThickness}Average stress at final temperature as a function of film thickness $H$ for $\phi = 30^\circ$ and $60^\circ$. The symbols represent results from the discrete dislocation simulations.}
\end{figure}

Fig.~\ref{fig:StressTemp} (a) and (b) show the average stress
$\stressAV$ as a function of temperature $T$ for different film
thicknesses for $\phi=30^\circ$ and $60^\circ$ respectively. When the
temperature axis is read right-to-left as a measure of strain, these
stress--strain curves are steeper (\emph{film is harder}) as the
thickness decreases. The hardening rate also increases with increasing
$\phi$, even though the Schmid factors for both orientations are
identical. Finally, the prediction of the average tensile stress versus film
thickness according to Eq.~(\ref{E:stressAV}) is shown in Fig.~\ref{fig:StressThickness} against the discrete dislocation results (in symbols) for both slip orientations with satisfactory agreement.

\begin{figure}[hbt!]
\psfrag{stress}{$\stressAV$ (MPa)}
\psfrag{thickness}{$H$ ($\mu$m)}
\psfrag{thirty}{$\phi = 30^\circ$}
\psfrag{sixty}{$\phi = 60^\circ$}
\psfrag{one}{$\propto 1/H$}
\psfrag{half}{$\propto 1/\sqrt{H}$}
\includegraphics[width=0.9\linewidth]{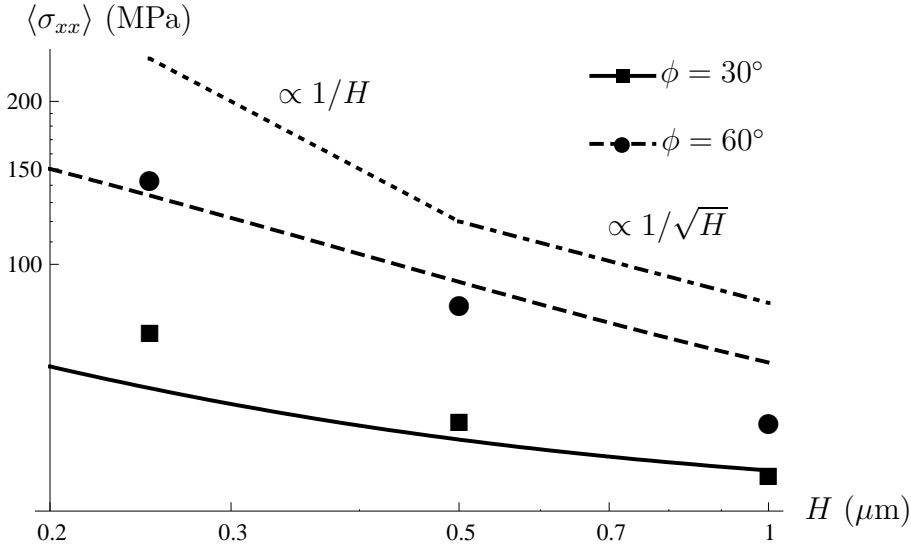}
\caption{\label{fig:StressThicknessLogLog}Log--log plot of the same
data as in Fig.~\ref{fig:StressThickness}, and comparison with simple
power law scaling laws with exponents $-0.5$ and $-1$.}
\end{figure}

The thickness dependence of stress predicted by Eq.~(\ref{E:stressAV})
is clearly a more complicated one than a simple scaling of the type
$\stressAV \propto H^{-p}$, with $p$ varying usually between $1/2$ and
$1$. In order to see how large this deviation is,
Fig.~\ref{fig:StressThicknessLogLog} shows the data of
Fig.~\ref{fig:StressThickness} on double-log scales.
For $\phi=60^\circ$, the theoretical $\stressAV(H)$ is rather close to
a power law over the entire regime considered here, but curves upwards
for very small $H$ when $\phi=30^\circ$. Enhanced hardening in very
thin films is observed in discrete dislocation results
\citep{NicoGiesNeed03,NicoGiesNeed05} and has been attributed there to
dislocation sources being shut down by relatively long pile-ups; this
effect is absent in the quasi-static solution developed here since
nucleation is not taken into account. 

A similar theoretical study has been carried out 
by \cite{NicoGiesGurt05} 
using Gurtin's strain-gradient theory. Compared to
the discrete dislocation results, size-dependent
hardening was captured but not the orientation dependence since Gurtin's original theory
predicts the same response for $\phi=30^\circ$ as for
$\phi=60^\circ$. Subsequently, they proposed a modified
``defect energy'' based on the
consideration of dislocation pile-ups which did predict the correct
$\phi$--trend. The latter implies a material length scale that scales with
$\cos{\phi}$, while our theory predicts scaling with $1/f(\phi) \propto
\sin^2{\phi}/\cos{\phi}$; the ratio of these for $\phi=30^\circ$ and
$60^\circ$ is identical. It is also interesting to note that the theory by
\cite{NicoGiesGurt05} reveals a constant hardening rate for a given thickness and slip orientation, whereas we find a weak logarithmic dependence on temperature. Both outcomes are within the error bar of the discrete dislocation results.

\section{Simple shear of constrained film}\label{S:Shearing}

\begin{figure}[htb]
\psfrag{s1}{$\hvect{s}_1$}
\psfrag{m1}{$\hvect{m}_1$}
\psfrag{s2}{$\hvect{s}_2$}
\psfrag{m2}{$\hvect{m}_2$}
\psfrag{f}{$\phi$}
\psfrag{x}{$\hvect{x}$}
\psfrag{y}{$\hvect{y}$}
\psfrag{H}{$H$}
\psfrag{G}{$\Gamma(t)$}
\includegraphics[width=0.8\linewidth]{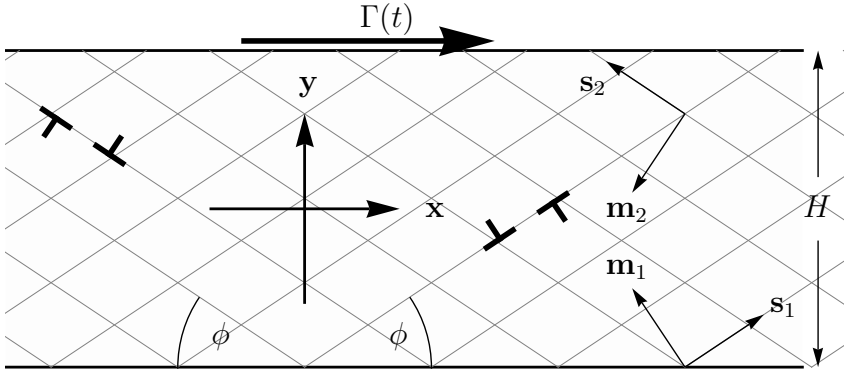}
\caption{\label{fig:Shear}The thin film of thickness $H$ with two impenetrable top and bottom layers is under prescribed shearing $\Gamma(t)$. The film has two symmetrical slip planes defined by angle $\phi$. The origin of the coordinate system is located at the center of the film.}
\end{figure}

We consider the same film as in the previous section, but now subjected to a
shear $\Gamma$ in the $\hvect{x}$ direction, see Fig.~\ref{fig:Shear}. 
While the normal strain was uniform in the
film under thermal straining, in the present problem the only
non-vanishing stress component $\sigma_{xy}$ is uniform across the width. The
second difference is that now both surfaces are impenetrable for dislocations;
i.e. $\gamma_1 = \gamma_2 = 0$ at $y = \pm H/2$ (note that 
the origin has been placed at the center of the film for calculational
convenience). By symmetry, $\tau_1 = \tau_2$ and $\gamma_1 = \gamma_2$ which implies that $\kappa_1 = \kappa_2$ and $\rho_1 = \rho_2$. We shall therefore omit the subscripts.

We can again solve this problem quasi-statically in the manner of Sec.~\ref{S:ThinFilm}. The resolved shear stress $\tau_1$ is given by
\begin{equation}
 	\tau = \cos(2\phi)\,\sigma_{xy},
\end{equation}
while the back stress $\tau^\text{b}$ is, according to Eq.~(\ref{E:taub}),
\begin{equation}
 	\tau^\text{b} = GDf(\phi)\,\frac{\partial_y\kappa}{\rho}\,,
\end{equation}
where $G \equiv \mu b/(2\pi(1-\nu))$ contains all the material parameters, and
$f(\phi) \equiv \sin(\phi)(1-\cos(2\phi))=2\sin^3(\phi)$ captures the slip
orientation information. Force balancing, $\tau - \tau^\text{b} = 0$, implies that
\begin{equation}
	\sigma_{xy} = GD\,\frac{f(\phi)}{\cos(2\phi)}\,\frac{\partial_y\kappa}{\rho}.
\end{equation}
Since $\sigma_{xy}$ is uniform across the film thickness by virtue of
equilibrium, we arrive at the differential equation
\begin{equation}\label{E:DE}
 	A(\phi)\partial_y \kappa(y) = \sigma_{xy}\rho(y)\,,
\end{equation}
above yield. Here $A(\phi) \equiv 2GD\sin^3(\phi)/\cos(2\phi)$ contains the slip orientation dependence.

Under shear, dislocations of one sign (negative when $\phi>\pi/4$) move towards the top
$y=H/2$ where they are blocked, while the opposite-signed dislocations will
pile-up against the bottom surface; this implies that $\kappa(y) = -\text{sign}(y)\rho(y)$.
The solution of Eq.~(\ref{E:DE}) is thus very simple:
\begin{equation}\label{E:kappaSoln}
 	\kappa(y) = -\text{sign}(y)\kappa_0\,\text{exp}\!\left[\sigma_{xy}\big|y/A(\phi)\big|\right]
\end{equation}
The constant of integration $\kappa_0$ in general could be a function of the
applied shear $\Gamma$. Using the relationship (\ref{E:Kroner}) between GND
density and slip, Eq.~(\ref{E:kappaSoln}) together with the no-slip boundary conditions give
\begin{equation}\label{E:gamma}
 	\gamma(y) = -\gamma_0(\Gamma)\left\{1 -\text{exp}\left[\lambda (|y|-H/2)\right]\right\},
\end{equation}
where all the integration constants have been absorbed into $\gamma_0(\Gamma)$,
and $1/\lambda \equiv \left|A(\phi)\right|/\sigma_{xy}(\Gamma)$ gives the
approximate characteristic width of the boundary layers as a function of the applied shear $\Gamma$.

Averaging of the decomposition $\varepsilon_{ij} = \varepsilon^\text{E}_{ij} +
\varepsilon^\text{P}_{ij}$ across the width of the sample, along with Hooke's
law $\sigma_{xy} = 2\mu \varepsilon^\text{E}_{xy}$ gives
\begin{equation}\label{E:decomp}
 	\sigma_{xy} = 2\mu\left(\Gamma/2 - \cos(2\phi)\avg{\gamma}\right).
\end{equation}
Here, we have made use of the fact that $\avg{\varepsilon_{xy}}=\Gamma/2$ and employed Eq.~(\ref{E:Orowan}) to find $\varepsilon^\text{P}_{xy} =
\cos(2\phi)\gamma$.
The average slip can be calculated directly from Eq.~(\ref{E:gamma}),
\begin{equation}\label{E:avggamma}
	\avg{\gamma} = -\gamma_0(\Gamma)\left[1-\frac{1-\exp{-\lambda H/2}}{\lambda H/2} \right].
\end{equation}
The functional form of $\gamma_0(\Gamma)$ can be obtained in the limit of large film thickness, $H\rightarrow \infty$, where the system is insensitive to the boundary layers which results in perfect plasticity. In this case Eq.~(\ref{E:decomp}) implies that
\begin{equation}\label{E:gamma0}
 	\tauY = \mu \left(\Gamma + 2\cos(2\phi)\gamma_0\right).
\end{equation}
Eqs.~(\ref{E:decomp})--(\ref{E:gamma0}) together provide an implicit expression of $\sigma_{xy}$ as a function of the applied shear $\Gamma$:
\begin{equation}\label{E:stressDep}
 	\sigma_{xy} = \tauY +(\mu\,\Gamma-\tauY)\,\frac{1-\exp{-\lambda H/2}}{\lambda H/2}
\end{equation}
\begin{figure}[htb]
\centering
\psfrag{yoH}{$y/H$}
\psfrag{twoeps}{$2\varepsilon_{xy}$}
\psfrag{Geq}{$\Gamma=$}
\includegraphics[width=0.8\linewidth]{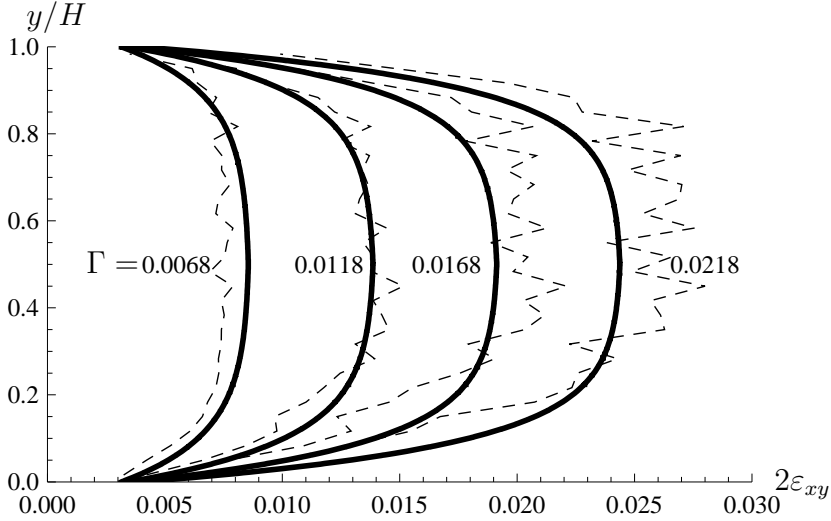}
\caption{\label{fig:yVSStrain}Comparison of the discrete dislocation~\citep{ShuFlecGiesNeed01} (dashed lines) and nonlocal plasticity (solid curves) shear strain profiles at different values of the applied shear $\Gamma$ for film thickness $H=1~\mu$m}
\end{figure}

\begin{figure}[htb]
\psfrag{stress}{$\sigma_{xy}/\sigma_0$}
\psfrag{Gamma}{$\Gamma$}
\psfrag{pfivemic}{$H=0.5\micron$}
\psfrag{onemic}{$1\micron$}
\psfrag{twomic}{$2\micron$}
\centering%
\includegraphics[width=0.8\linewidth]{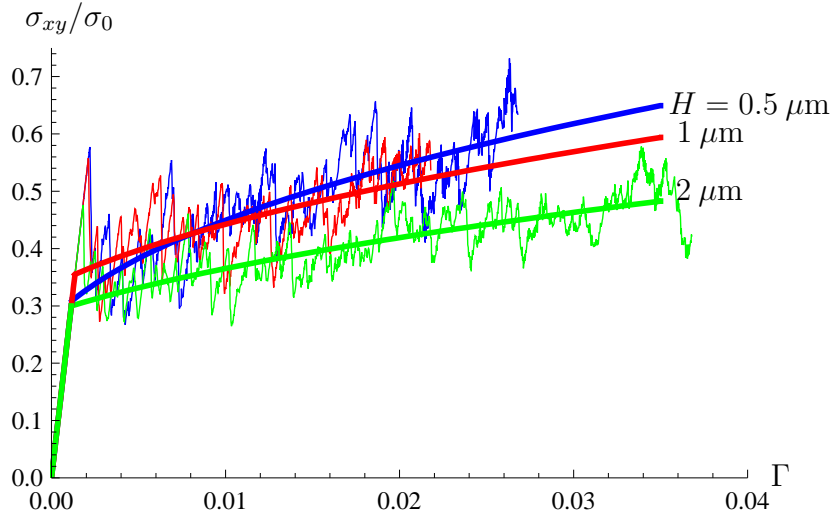}
\caption{\label{fig:StressVSGamma}Shear stress response to applied shear
  $\Gamma$ for various film thicknesses. The discrete dislocation data (dashed
  lines) is taken from \cite{ShuFlecGiesNeed01}.}
\end{figure}

The continuum theory is tested against the discrete dislocation
simulations by \cite{ShuFlecGiesNeed01} on a crystal with two slip systems
oriented at $\phi = 60^\circ$. The elastic properties are the same as in
Sec.~\ref{S:DFT}, i.e. $\mu = 26.3$~GPa and $\nu = 0.33$, and stress is
measured in units of the mean nucleation strength $\sigma_0 = 1.9\times
10^{-3}\mu$ in the discrete simulations. We first note that the width of the
boundary layers $1/\lambda$ cannot be used as a fitting parameter since its
value changes with increasing stress. We therefore define the length parameter
$l \equiv \sigma_{xy}/(\sigma_0\lambda) = |A|/\sigma_0$, which is independent
of $\sigma_{xy}$, as a new fitting parameter. Given stress $\sigma_{xy}$ at a
selected shear $\Gamma$, the value of $l$ can be determined from fitting
Eq.~(\ref{E:gamma}) to the strain distribution across the film thickness, as
shown in Fig.~\ref{fig:yVSStrain}.
The fitting procedure is somewhat intricate due to the non-algebraic nature of
Eq.~(\ref{E:stressDep}) which needs to be computed for $\gamma(y)$ in
Eq.~(\ref{E:gamma}) at a given $\Gamma$. We therefore take the stress value
from the simulation stress--shear curve (Fig.~\ref{fig:StressVSGamma}) as an
additional input for the fitting of $l$, yielding $l \simeq 46$~nm for the
case of $H = 1~\mu$m on the basis of the stress at $\Gamma = 0.0218$. Fig.~\ref{fig:yVSStrain} shows shear strain distributions across the film thickness at three other values of $\Gamma$ where no additional fitting has been performed.

\begin{figure}[b!]
\psfrag{yoH}{$y/H$}
\psfrag{kappa}{$\kappa~(\micron^{-2})$}
\centering%
\includegraphics[width=\linewidth]{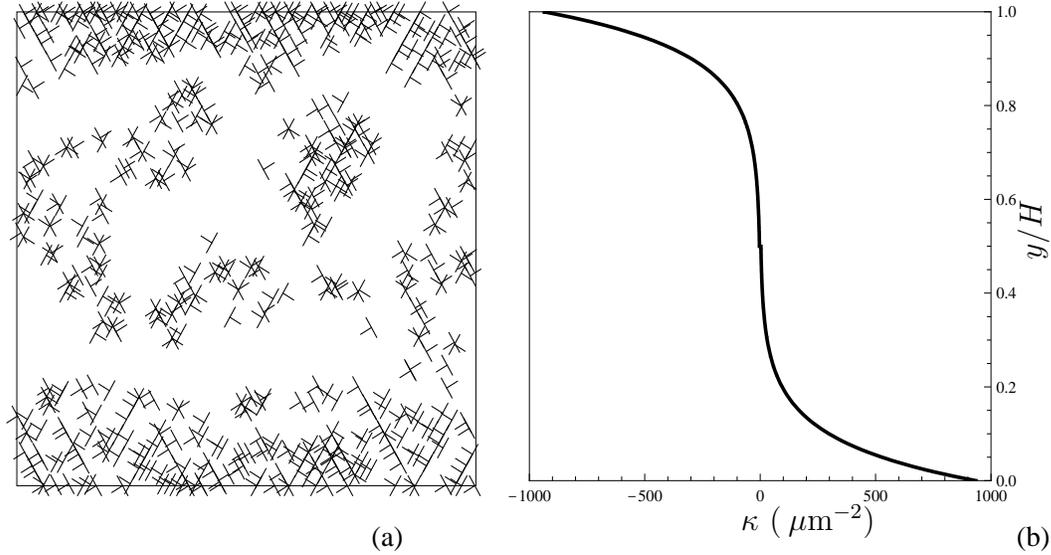}
 \caption{\label{fig:DislProfiles60}Dislocation distribution in a $H=1\micron$
   thick $\phi=60^\circ$ film at an overall shear of $\Gamma=0.0218$ according to
   (a) discrete simulations by \cite{ShuFlecGiesNeed01} and (b) the theoretical
   solution (\ref{E:kappaSoln}) with $l \simeq 46$~nm.}
\end{figure}

For further comparison, Fig.~\ref{fig:DislProfiles60} (b) shows the
theoretical distribution of dislocation density (recall that $\rho=\kappa$ for
all slip systems) in comparison with the discrete dislocation distribution in
a periodic cell with a width of $1\micron$. The theory correctly predicts the
development of intense dislocation boundary layers. The core of the crystal is
left almost dislocation free as dislocations pass each other almost unhindered
on their way towards the top and bottom faces.

From the above-mentioned best-fit $l$ for the film thickness of $H=1\micron$,
we can study the shear response for different film thicknesses. Data of the
discrete dislocation simulations suggest thickness-dependent initial yield
strengths. The responses are shown in Fig.~\ref{fig:StressVSGamma} in
comparison with results from the discrete simulations. We supply for each film
thickness the best-fit yield point as an extra degree of freedom. Similar to
the previous test problem (Sec.~\ref{S:ThinFilm}), the stress-strain curves
show size-dependent hardening. The hardening rate decreases with increasing
applied external shear, and approaches a constant value at large
shear. \cite{ShuFlecGiesNeed01} also analyzed this problem with their version
of strain-gradient theory and found weak size effects. Their stress response,
however, is linear due to the fact that the width of dislocation boundary
layers is constant in their theory. The same linear stress-strain relation was
also predicted by Gurtin's strain-gradient theory \citep{BittNeedGurtGies03}.

It should be mentioned that the exact form of the slip-interaction coupling (in Eq.~(\ref{E:taub})) turns out to be unimportant in this problem. The slip orientation dependence is buried inside the definition of $|A(\phi)|$ which has been absorbed into the fitting parameter $l$. On this ground, it does not matter whether this coupling be $\cos(\theta_{ij})$ or $\cos(2\theta_{ij})$ as proposed by \cite{YefiGies05}. 

\begin{figure}
\psfrag{yoH}{$y/H$}
\psfrag{kappa}{$\kappa~(\micron^{-2})$}
\includegraphics[width=\linewidth]{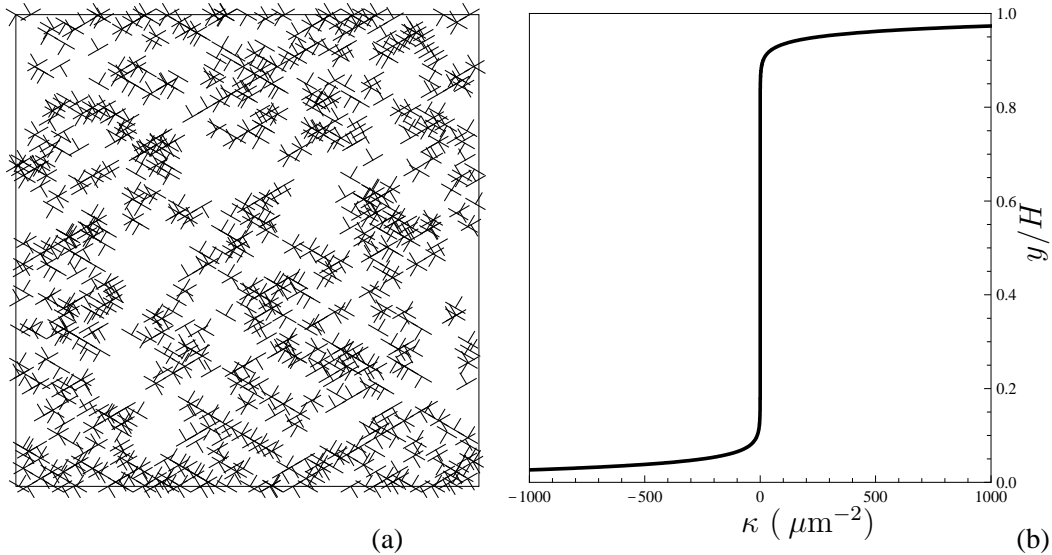}
 \caption{\label{fig:DislProfiles30}Dislocation distribution in a $H=1\micron$
   thick $\phi=30^\circ$ film at an overall shear of $\Gamma=0.0218$ according to
   (a) discrete simulations by \cite{ShuFlecGiesNeed01} and (b) the theoretical
   solution (\ref{E:kappaSoln}) with $l \simeq 8.7$~nm.}
\end{figure}
Our theory predicts drastic changes in behavior when $\phi$ crosses
$45^\circ$. Due to a sign change in the resolved shear stress, the charges of
dislocations at the two interfaces reverse from the present situation when
$\phi < 45^\circ$ (resulting in the sign alternations of $\kappa_0$ and
$\gamma_0$ in Eqs.~(\ref{E:kappaSoln}), (\ref{E:gamma}), (\ref{E:avggamma}),
and (\ref{E:gamma0})). As a result, the applied shear acts in favor of the new
dislocation arrangement---in other words---our theory predicts that the back
stress further enhances plasticity instead of impeding the flow of new
dislocations into the boundaries. Hence, thinner boundary layers are expected
which suggests smaller size effects. More quantitatively, for the orientation
angle of, say, $\phi = 30^\circ$, the layers should be thinner by a factor of
$l_{30^\circ}/l_{60^\circ} =
|A({30^\circ})/A({60^\circ})| = (\sin(30^\circ)/\sin(60^\circ))^3 \simeq
0.19$. The dislocation distribution thus predicted is shown in
Fig.~\ref{fig:DislProfiles30}(b). 

\begin{figure}
 \begin{minipage}{0.5\linewidth}
   \centering
   \includegraphics[width=0.9\linewidth]{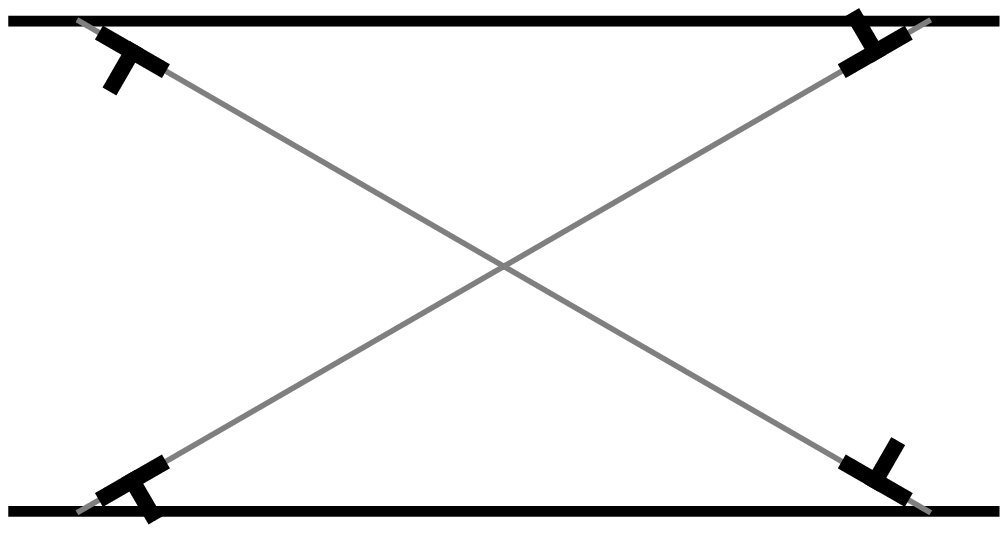}\\
   (a)
 \end{minipage}%
 \begin{minipage}{0.5\linewidth}
   \centering
   \includegraphics[width=0.9\linewidth]{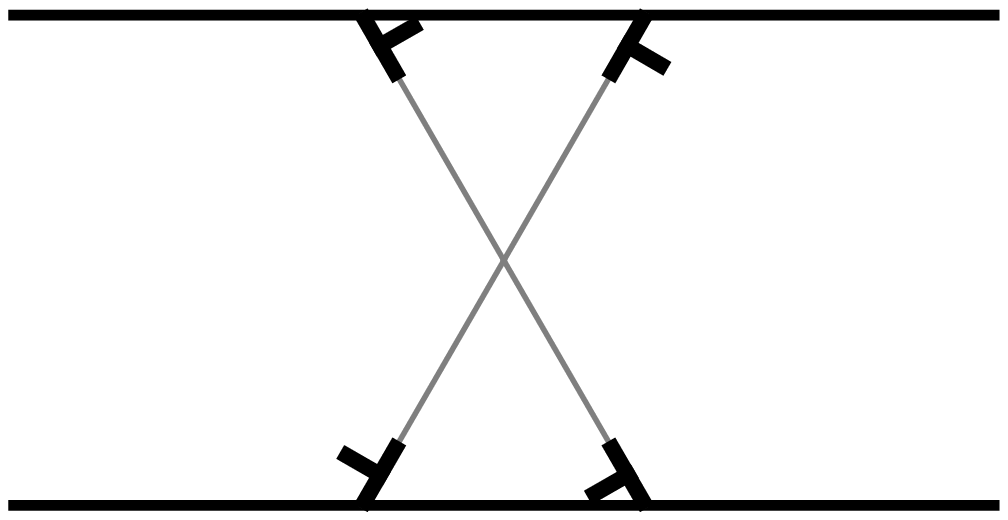}\\
   (b)
 \end{minipage}\\
 \begin{minipage}{0.5\linewidth}
   \centering
   \includegraphics[width=0.7\linewidth]{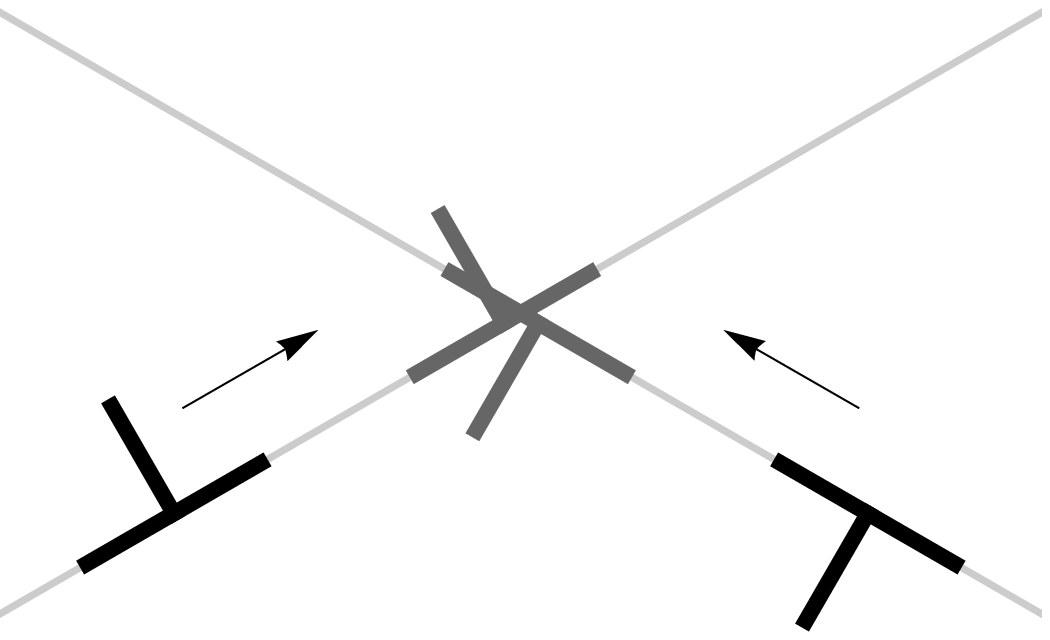}\\
   (c)
 \end{minipage}%
 \begin{minipage}{0.5\linewidth}
   \centering
   \includegraphics[width=0.7\linewidth]{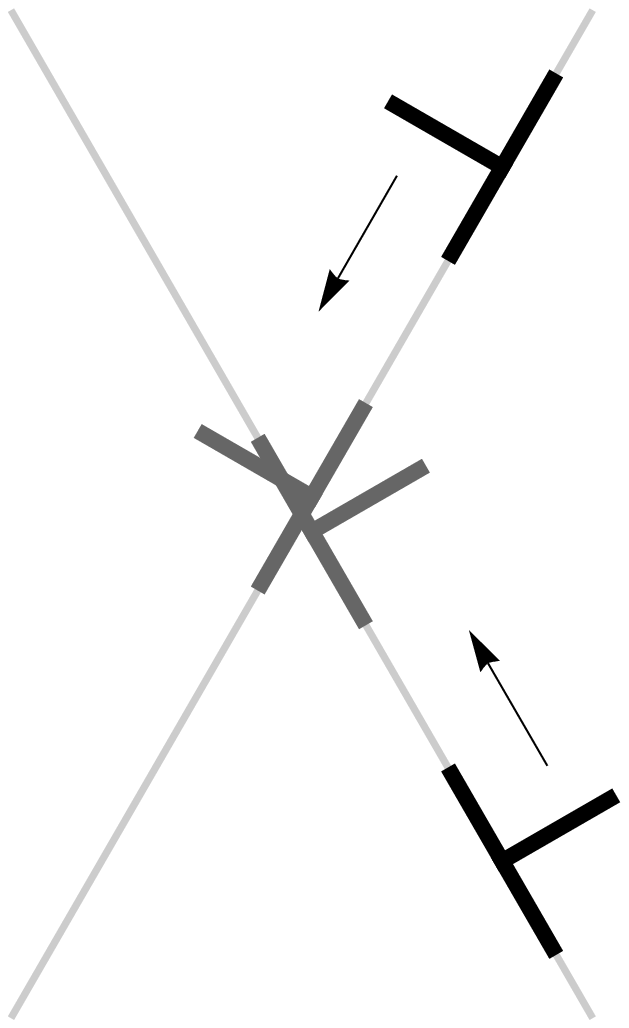}\\
   (d)
 \end{minipage}
\caption{\label{fig:Locking}Types of dislocations at the pile-ups for (a) $\phi=30^\circ$ and (b) $\phi=60^\circ$, and their locking situations as they glide (a) upwards or downwards (as viewed upwide down) for $30^\circ$ case or (d) in the opposite directions for $60^\circ$ case.}
\end{figure}
Discrete dislocation dynamics simulations, however, reveal essentially no
boundary layers at all---or, equivalently, boundary layers that span the
entire width (Fig.~\ref{fig:DislProfiles30}(a)). Upon closer
examination, we find `locks' of dislocations\footnote{The use of the phrase
  `locks' for parallel edge
  dislocations is somewhat inappropriate, because such dislocations only
  interact through their long-range field but do not alter the topology as
  happens in, e.g., Lomer locks. We nevertheless use the term here because it
  expresses the small-scale interactions that obstruct dislocation
  motion.} 
of like charges on different
slip systems which prevent their motion pile-ups to the boundaries. A pair of
dislocations with the relative angle of their Burgers vectors between
$90^\circ$ and $270^\circ$ feel their mutual attraction when they glide past
each other. Although rather weak, this interaction is apparently strong enough
in this case for locking to occur. Figs.~\ref{fig:Locking}(a) and (b) show the types of dislocations which accumulate at the boundaries for $\phi = 30^\circ$ and $60^\circ$, respectively. Figs.~\ref{fig:Locking}(c) and (d) demonstrate one of the two situations when locking happens in each case (the others are $180^\circ$ rotations of these). In the region sufficiently far away from the boundaries, event \ref{fig:Locking}(c) is roughly as likely
to occur as event \ref{fig:Locking}(d), since the situations differ just by a
$90^\circ$ rotation followed by a flip about the $\hvect{y}$-axis. The relative likelihood, however,
increases immensely close to the boundaries because only in the $30^\circ$
case do dislocations moving to the same boundary permit locking,
Fig.~\ref{fig:Locking}(c). This mutual locking of slip systems prevents
dislocations to reach the boundaries and form localized boundary layers. The locking mechanism is purely a discrete phenomena and cannot be captured by the current continuum theory without further refinement. Due to its relatively small probability, locking seldom occurs in the $60^\circ$ case.

\section{Discussion and Conclusion}
We applied the recently formulated multislip continuum plasticity theory to
analyze two boundary value problems relating to thin films. In
Sec.~\ref{S:ThinFilm}, we studied stress relaxation mechanism in thin films on
substrates with thermal loading. We obtained an explicit analytical expression
of the stress distribution as a function of slip orientation with one fitting
parameter. The predictions were in good agreement with the discrete
dislocation results of \cite{NicoGiesNeed03,NicoGiesNeed05}. Our theory was
able to show size-dependent hardening and the hardening due to slip
orientations---both of which the previous continuum theory failed to
explain. Subsequently, we analyzed simple shear in constrained
films. Similarly to the first problem, we observed dislocation pile-ups at the
top and bottom constrained boundaries. The thickness of dislocation layers
depends weakly on the incremental shear. Regardless of the difference between the
forms of slip-interaction coupling between our theory and that in
\cite{YefiGies05}, our theory also gave satisfactory agreements with results
from discrete dislocation dynamics simulations \citep{ShuFlecGiesNeed01}. We
pointed out that this term can be absorbed into fitting parameter; the correct functional form of the coupling, therefore, cannot be decided only on the basis of this problem.

\section*{Acknowledgments}
We acknowledge funding from the European Commissions
Human Potential Programme {\textsc SizeDepEn| under contract number MRTN-CT-2003-504634.

\bibliographystyle{elsart-harv}
\bibliography{references}

\end{document}